\documentclass[times,doublespace]{rncauth}
\usepackage{moreverb}
\usepackage{rawfonts,graphicx,amsfonts,amssymb,psfrag,flushend}

\graphicspath{{simulation_graph/}}

\newtheorem{thm}{Theorem}

\newtheorem{lem}{Lemma}

\newtheorem{df}{Definition}

\begin{document}

\title{Robust time-varying formation design for multi-agent systems with disturbances: Extended-state-observer method}

\author{Le~Wang,~Jianxiang~Xi$^{*}$,~Ming~He,~Guangbin~Liu}
\address{\hspace{10.8em}Rocket Force University of Engineering, Xi'an, 710025, P.R. China\\}

\cgsn{This work was supported by the National Natural Science Foundation of China under Grants  61867005, 61763040, 61703411 and 61374054, Innovation Foundation of High-Tech Institute of Xi'an (2015ZZDJJ03) and Scientific Research Project under Grant JJ20172B03058, also supported by Innovation Zone Project under Grant 18-163-11-ZT-004-005-01. }{}

\begin{abstract}
~~Robust time-varying formation design problems for second-order multi-agent systems subjected to external disturbances are investigated. Firstly, by constructing an extended state observer, the disturbance compensation is estimated, which is a critical term in the proposed robust time-varying formation control protocol. Then, an explicit expression of the formation center function is determined and impacts of disturbance compensations on the formation center function are presented. With the formation feasibility conditions, robust time-varying formation design criteria are derived to determine the gain matrix of the formation control protocol by utilizing the algebraic Riccati equation technique. Furthermore, the tracking performance and the robustness property of multi-agent systems are analyzed. Finally, the numerical simulation is provided to illustrate the effectiveness of theoretical results.\par
\end{abstract}


\maketitle \vspace{-2em}

\section{Introduction}\label{section1} \vspace{-1em}

~~Recently, the distributed cooperative control of multi-agent systems has aroused extensive attentions in multiple fields, including wireless communication, robotics and distributed computation as shown in \cite{c1}-\cite{c8}. As an important research topic on the distributed control, formation control refers to design a control strategy with the neighboring information such that a group of autonomous agents reach an expected geometrical shape. In the past two decades, several classical formation control methodologies were investigated, such as the leader-follower method \cite{c9}, the virtual-structure-based strategy \cite{c10} and the behavioral approach \cite{c11}, among many others. Beard {\it et al}. \cite{c12} showed that each of the above-mentioned classical methodologies has its corresponding weakness. Ren \cite{c13} addressed formation control problems by implementing the consensus-based approach and showed that the above-mentioned classical methodologies could be unified in the framework of the consensus-based formation control. Inspired by the development of the consensus theory in the control community (e.g., \cite{c14}-\cite{c19}), the newly developed consensus-based formation control strategies were reported in many application fields including mobile robots, intelligent ground vehicles and unmanned aerial vehicles (see \cite{c20}-\cite{c25} and the references therein). \par

In many practical circumstances, multi-agent systems may suffer external disturbances due to environmental uncertainties, which can drive these systems to oscillations or divergences. For example, in the formation flying of multiple quadrotors, atmospheric disturbances can be regarded as additional forces and may cause instabilities in both the attitude and position dynamics. It is significant to address disturbance rejection problems such that multi-agent systems can reach the asymptotical disturbance rejection while conserving the closed-loop stability. Jafarian {\it et al}. \cite{c26} studied the formation keeping control for a group of nonholonomic wheeled robots with matched input disturbances, where the disturbances were compensated by internal-model-based controllers. In \cite{c27}, the time-invariant formation tracking control for a group of quadrotors with unknown bounded disturbances was achieved by designing an ${H_\infty }$ control controller, where the disturbance cannot be rejected by the proposed method in the whole desired frequency range. Liu {\it et al}. \cite{c28} proposed a robust compensating filter to handle time-invariant formation control problems of multiple quadrotors with disturbance rejections in the whole frequency domain as much as desired. \par

Note that the desired formation was time-invariant in \cite{c20}-\cite{c28}. However, time-varying formation configurations are required in many applications due to complex external environments and/or variable mission situations. For example, the formation shape should be changed in the obstacle avoidance for multiple mobile robots. Several significant results about the time-varying formation control were obtained in \cite{c29}-\cite{c33}. Cooperative time-varying formation control methods was studied in \cite{c29}, where the formation was characterized by time-varying external parameters. A time-varying formation of collaborative unmanned aerial vehicles and unmanned ground vehicles was achieved in \cite{c30}. Time-varying formation tracking control was reached when considering the influence of switching topologies in \cite{c31}. Dong {\it et al}. \cite{c32} investigated the time-varying formation analysis and design problems for second-order multi-agent systems with directed topologies, where a formation feasibility condition was proposed to show that not all expected formation could be achieved. Time-varying group formation control for multi-agent systems with directed topologies was showed in \cite{c33}. However, further investigates on disturbance rejections of the time-varying formation with the influence of external disturbances were not considered in \cite{c29}-\cite{c33}, and the disturbance rejection methods for time-invariant formations in \cite{c26}-\cite{c28} cannot be implemented since the expected formation is time-varying. To the best of our knowledge, robust time-varying formation design problems for second-order multi-agent systems with unknown external disturbances have not been investigated extensively. \par

Motivated by the above-mentioned facts, the current paper develops an extended-state-observer method to tackle the robust time-varying formation control problem for multi-agent systems subjected to external disturbances. With the disturbance compensation, a novel robust time-varying formation control protocol is proposed, using only relative neighboring information. By regarding external disturbances as additional states, an extended state observer (ESO) is constructed to determine the disturbance compensation. Then, the closed-loop dynamics of the whole multi-agent system is divided into two parts. The first one is the formation agreement dynamics, which is utilized to derive an explicit expression of the formation center function. The second part, called the disagreement dynamics, can describe the relative motion among agents. Sufficient conditions of the robust time-varying formation design are determined via algebraic Riccati equation techniques, together with the formation feasibility conditions. Moreover, the tracking performance and the robustness stability of the closed-loop system are analyzed. \par

Compared with the existing results about the time-varying formation control of multi-agent systems, the current paper contains the following three novel features. Firstly, to achieve the disturbance rejection control objective, a robust time-varying formation control protocol is proposed with the robust disturbance compensation. Time-varying formation protocols in \cite{c29}-\cite{c33} cannot deal with the robust time-varying formation control problems when the influence of external disturbances is considered. Secondly, an ESO is constructed to determine the robust disturbance compensation, which can actively compensate the external disturbance in real time. Tracking performances and robust properties are analyzed with the ESO and the formation feasibility condition. However, disturbance compensations and robust properties were not considered in \cite{c29}-\cite{c33}. Thirdly, an explicit expression of the formation center function is deduced to show the macroscopic motion of the whole formation under the influence of external disturbances. It is revealed that that the disturbance compensation has effects on formation center functions. However, \cite{c29}-\cite{c31} did not give the formation center function and the formation center functions in \cite{c32} and \cite{c33} could not determine the impact of disturbance compensation. \par

An outline of the current paper is presented as follows. Section \ref{section2} gives the problem description. In Section \ref{section3}, an explicit expression of the formation center functions is determined. In Section \ref{section4}, sufficient conditions of the robust time-varying formation design are shown and the tracking performance and the robustness stability are analyzed. Section \ref{section5} illustrates the effectiveness of theoretical results via a numerical simulation. Conclusions are stated in Section \ref{section6}.\par

Notations: Let ${\mathbb{R}^n}$ and ${\mathbb{R}^{n \times m}}$ be the $n$-dimension real column vector and the $n \times m$-dimension real matrix, respectively. For simplicity, $0$ uniformly represents the zero number, zero vectors and zero matrices. ${{\mathbf{1}}_N}$ stands for an $N$-dimensional column vector with each entry being $1$. ${P^{ - 1}}$, ${P^H}$ and ${P^T}$ denote the inverse matrix, the Hermitian adjoint matrix and the transpose matrix of $P$, respectively. The norm used here is respectively defined as ${\left\| {h(t)} \right\|_1} = {\max _i}\left( {\sum\nolimits_j {\int_0^\infty  {\left| {{h_{ij}}(t)} \right|dt} } } \right)$, $\left\| {p(t)} \right\| = {\left\| {p(t)} \right\|_2} = {\left( {\sum\nolimits_{i = 1}^n {{{\left| {{p_i}(t)} \right|}^2}} } \right)^{{1 \mathord{\left/ {\vphantom {1 2}} \right. \kern-\nulldelimiterspace} 2}}}$ and ${\left\| {p(t)} \right\|_\infty } = {\max _i}{\sup _{t \geqslant 0}}\left| {{p_i}(t)} \right|$, where $\left|  \cdot  \right|$ is the absolute value, $h(t) = \left[ {{h_{ij}}(t)} \right] \in {\mathbb{R}^{m \times n}}$ and $p(t) = \left[ {{p_i}(t)} \right] \in {\mathbb{R}^n}$. The Kronecker product is represented by the notation $ \otimes $.\par

\section{Problem description }\label{section2} \vspace{-0.5em}

~~~~Consider a group of $N$ identical agents with the dynamics of the $i$th agent described by:
\begin{eqnarray}\label{1}
\left\{ {\begin{array}{*{20}{c}}
   {{{\dot p}_i}(t) = {v_i}(t),} \hfill  \\
   {{{\dot v}_i}(t) = {\alpha _p}{p_i}(t) + {\alpha _v}{v_i}(t) + {u_i}(t) + {\omega _i}(t),} \hfill  \\
 \end{array} } \right.
\end{eqnarray}
where $i = 1,2, \cdots ,N $, ${p_i}(t)$, ${v_i}(t)$ and ${u_i}(t) \in {\mathbb{R}^n}$ represent the position, the velocity and the control input, respectively, ${\omega _i}(t) \in {\mathbb{R}^n}$ is the unknown bounded external disturbance and ${\alpha _p}$ and ${\alpha _v}$ are the damping constants. The interaction topology among agents is described by a digraph $G$, where agent $i$ is represented by the  $i$th node, the interaction channel among nodes is denoted by an edge and the interaction strength is depicted by the edge weight ${w_{ij}}$. Note that ${w_{ij}} > 0$ if agent $j$ belongs to the neighbor set ${N_i}$ of agent $i$ and ${w_{ij}} = 0$ otherwise. For the digraph $G$, the weighted adjacency matrix is $W = {[{w_{ij}}]_{N \times N}}$ and $D = {\text{diag}}\{ {d_1},{d_2}, \cdots ,{d_N}\} $ stands for the in-degree matrix. Define the Laplacian matrix of $G$ as $L = D - W$. A directed path from node $i$ to node $j$ is a finite ordered sequence of edges described as $\left\{ {({v_i},{v_m}),({v_m},{v_n}), \cdots ,({v_l},{v_j})} \right\}$. A digraph is said to have a spanning tree if a root node $i$ exists such that it at least has a directed path to every other node.

\begin{lem} [\cite{c34}]\label{lemma1}
If $G$ has a spanning tree, then $0$ is its single eigenvalue with ${\mathbf{1}_N}$ being the related eigenvector and other $N - 1$ eigenvalues have positive real parts; that is, $0 = {\lambda _1} < \operatorname{Re} ({\lambda _2}) \leqslant  \cdots  \leqslant \operatorname{Re} ({\lambda _N})$.
\end{lem}

Let ${f_i}(t) = {[f_{ip}^T(t),f_{iv}^T(t)]^T} \in {\mathbb{R}^{2n}}$ $(i \in \{ 1,2, \cdots ,N\} )$ be a piecewise continuously differentiable vector, then the expected time-varying formation is specified by a vector $f(t) = {[f_1^T(t),f_2^T(t), \cdots ,f_N^T(t)]^T}$. By considering the disturbance compensation, a robust time-varying formation control protocol is proposed as follows:
\begin{eqnarray}\label{2}
{u_i}(t) = {K_u}\sum\limits_{j \in {N_i}} {{w_{ij}}\left( {{x_j}(t) - {x_i}(t) - {f_j}(t) + {f_i}(t)} \right)}  - \alpha {f_i}(t) + {\dot f_{iv}}(t) - {z_i}(t),
\end{eqnarray}
where $i \in \{ 1,2, \cdots ,N\} $, ${x_i}(t) = {[p_i^T(t),v_i^T(t)]^T}$, $\alpha  = [{\alpha _p},{\alpha _v}] \otimes {I_n}$, ${K_u} \in {\mathbb{R}^{n \times 2n}}$ is the gain matrix and ${z_i}(t)$ is the robust disturbance compensation, which is determined by the following ESO:
\begin{eqnarray}\label{3}
\left\{ \begin{gathered}
  {{\dot g}_i}(t) = {z_i}(t) + {u_i}(t) + {\alpha _p}{p_i}(t) + {\alpha _v}{v_i}(t) - {\beta _{ig}}\left( {{g_i}(t) - {v_i}(t)} \right), \hfill \\
  {{\dot z}_i}(t) =  - {\beta _{iz}}\left( {{g_i}(t) - {v_i}(t)} \right), \hfill \\
\end{gathered}  \right.
\end{eqnarray}
where ${\beta _{ig}}$ and ${\beta _{iz}}$ are bandwidth constants, and ${g_i}(t)$ is the intermediate variable of the ESO.\par

Let $x(t) = {[x_1^T(t),x_2^T(t), \cdots ,x_N^T(t)]^T}$, ${\theta _1} = {[1,0]^T} \otimes {I_n}$ and ${\theta _2} = {[0,1]^T} \otimes {I_n}$, then multi-agent system (\ref{1}) with protocol (\ref{2}) can be rewritten as a global closed-loop system with the following dynamics:
\[  \dot x(t) = \left( {{I_N} \otimes \left( {{\theta _1}\theta _2^T + {\theta _2}\alpha } \right) - L \otimes {\theta _2}{K_u}} \right){x}(t) - \left( {{I_N} \otimes {\theta _2}\alpha  - L \otimes {\theta _2}{K_u}} \right)f(t)\]
\vspace{-3em}
\begin{eqnarray}\label{4}
\hspace{6em} + \left( {{I_N} \otimes {\theta _2}\theta _2^T} \right)\dot f(t) + \left( {{I_N} \otimes {\theta _2}} \right)\left( {\omega (t) - z(t)} \right).
\end{eqnarray}

\begin{df} \label{definition1}
For any given positive constant $\varepsilon $ and bounded initial states $x(0)$, multi-agent systems (\ref{1}) is said to be robust time-varying formation-reachable by protocol (\ref{2}) if all states involved in the global closed-loop system (\ref{4}) are bounded and there exist a gain matrix ${K_u}$, a vector-valued function $c(t)$ and a finite constant ${t_\varepsilon }$ such that $\left\| {{x_i}(t) - {f_i}(t) - c(t)} \right\| \leqslant \varepsilon $ $(\forall i \in \{ 1,2, \cdots ,N\} )$, $\forall t \geqslant {t_\varepsilon}$, where $c(t)$ is the formation center function and $\varepsilon $ is called the time-varying formation error bound, respectively.\par
\end{df}

The control objective of the current paper is to design the robust time-varying formation control protocol such that second-order multi-agent systems with external disturbances can reach the expected robust time-varying formation. The following three problems are focused: (i) Determining an explicit expression of formation center functions; (ii) Designing the gain matrix ${K_u}$ of protocol (\ref{2}); (iii) Analyzing the tracking performance and the robustness property of the global closed-loop system.\par

\section{Formation center functions}\label{section3}\vspace{-0.5em}

~~~This section gives an explicit expression of formation center functions and shows impacts of the time-varying formation and the disturbance compensation on the formation center function, respectively. \par

Let ${\xi _i}(t) = {x_i}(t) - {f_i}(t)$ $(i \in \{ 1,2, \cdots ,N\} )$ and $\xi (t) = {[\xi _1^T(t),\xi _2^T(t), \cdots ,\xi _N^T(t)]^T}$, then global closed-loop system (\ref{4}) can be transformed into
\[  \dot \xi (t) = \left( {{I_N} \otimes \left( {{\theta _1}\theta _2^T + {\theta _2}\alpha } \right) - L \otimes {\theta _2}{K_u}} \right)\xi (t) + \left( {{I_N} \otimes {\theta _2}} \right)\left( {{\omega}(t) - {z}(t)} \right) \]
\vspace{-3em}
\begin{eqnarray}\label{5}
  \hspace{6em} + \left( {{I_N} \otimes {\theta _1}\theta _2^T} \right)f(t) - \left( {{I_N} \otimes {\theta _1}\theta _1^T} \right)\dot f(t).
\end{eqnarray}
Let $U = [{{\mathbf{1}}_N},\tilde u] \in {\mathbb{R}^{N \times N}}$ be a nonsingular matrix with $\tilde u = [{\tilde u_2},{\tilde u_3}, \cdots ,{\tilde u_N}] \in {\mathbb{R}^{N \times (N - 1)}}$ such that ${U^{ - 1}}LU = J$, where ${U^{ - 1}} = {[\bar u_1^H,{\bar u^H}]^H}$ with $\bar u = {[\bar u_2^H,\bar u_3^H, \cdots ,\bar u_N^H]^H} \in {\mathbb{R}^{(N - 1) \times N}}$ and $J$ is the Jordan canonical form of $L$.\par

According to Lemma \ref{lemma1} and the structure of $U$, one can obtain that $J = \text{diag}\{ 0,\tilde J\} $, where $\tilde J$ consists of the corresponding Jordan blocks of ${\lambda _i}$ $(i = 2,3, \cdots ,N)$. Let $\tilde \xi (t) = ({U^{ - 1}} \otimes {I_{2n}})\xi (t) = {[{\kappa ^T}(t),{\varphi ^T}(t)]^T}$, in which $\kappa (t) = {\tilde \xi _1}(t)$ and $\varphi (t) = {[\tilde \xi _2^T(t),\tilde \xi _3^T(t), \cdots ,\tilde \xi _N^T(t)]^T}$, then multi-agent system (\ref{5}) can be transformed into
\begin{eqnarray}\label{6}
\dot \kappa (t) = \left( {{\theta _1}\theta _2^T + {\theta _2}\alpha } \right)\kappa (t) + \left( {{{\bar u}_1} \otimes {\theta _2}} \right)\left( {\omega (t) - z(t)} \right) + \left( {{{\bar u}_1} \otimes {\theta _1}} \right)\left( {{f_v}(t) - {{\dot f}_p}(t)} \right),
\end{eqnarray}
\[  \dot \varphi (t) = \left( {{I_{N - 1}} \otimes \left( {{\theta _1}\theta _2^T + {\theta _2}\alpha } \right) - \tilde J \otimes {\theta _2}{K_u}} \right)\varphi (t) + (\bar u \otimes {\theta _2})\left( {\omega (t) - z(t)} \right) \]
\vspace{-3em}
\begin{eqnarray}\label{7}
  \hspace{8em} + \left( {\bar u \otimes {\theta _1}} \right)\left( {{f_v}(t) - {{\dot f}_p}(t)} \right),
\end{eqnarray}
where ${f_v}(t) = {[f_{1v}^T(t),f_{2v}^T(t), \cdots ,f_{Nv}^T(t)]^T}$ and ${\dot f_p}(t) = {[\dot f_{1p}^T(t),\dot f_{2p}^T(t), \cdots ,\dot f_{Np}^T(t)]^T}$. \par

Subsystems (\ref{6}) and (\ref{7}) depict the formation agreement and disagreement dynamics of multi-agent system (\ref{1}), which describe the absolute movement of the whole system and the relative movement among agents, respectively. According to subsystem (\ref{6}), the following theorem determines the impact of the disturbance compensation on the formation center function and shows an explicit expression of the formation center function, which describes the macroscopic motion of the whole formation.\par

\begin{thm} \label{theorem1}
For any given $\varepsilon  > 0$, if multi-agent system (\ref{1}) reaches the expected robust time-varying formation $f(t)$, then the formation center function $c(t)$ satisfies that
\[\left\| {c(t) - {c_0}(t) - {c_z}(t) - {c_f}(t)} \right\| \leqslant \varepsilon ,{\text{ }}\forall t \geqslant {{t_\varepsilon }},\]
where ${t_\varepsilon }$ is a finite constant and \vspace{-5pt}
\[{c_0}(t) = {e^{({\theta _1}\theta _2^T + {\theta _2}\alpha )t}}({\bar u_1} \otimes {I_{2n}})x(0),\]
\[{c_z}(t) = \int_0^t {{e^{({\theta _1}\theta _2^T + {\theta _2}\alpha )(t - \varsigma )}}\left( {{{\bar u}_1} \otimes {\theta _2}} \right)\left( {\omega (\varsigma ) - z(\varsigma )} \right)} ds,\]
\[{c_f}(t) = \int_0^t {{e^{({\theta _1}\theta _2^T + {\theta _2}\alpha )(t - \tau )}}\left( {{{\bar u}_1} \otimes {\theta _2}} \right)\left( {{{\dot f}_v}(\tau ) - {\alpha _p}{f_p}(\tau ) - {\alpha _v}{f_v}(\tau )} \right)} d\tau  - \left( {{{\bar u}_1} \otimes {I_{2n}}} \right)f(t).\]
\end{thm}\vspace{-10pt}

\begin{proof}
Let ${e_1} \in {\mathbb{R}^N}$ denote a unit vector with its first element being $1$. Define the following auxiliary functions:
\begin{eqnarray}\label{8}
{\xi _a}(t) = (U \otimes {I_{2n}}){[{\kappa ^T}(t),0]^T},
\end{eqnarray}
\begin{eqnarray}\label{9}
{\xi _d}(t) = (U \otimes {I_{2n}}){[0,{\varphi ^T}(t)]^T},
\end{eqnarray}
with $\left\| {U \otimes {I_{2n}}} \right\| = {\varepsilon _N}$. Due to $({U^{ - 1}} \otimes {I_{2n}})\xi (t) = {[{\kappa ^T}(t),{\varphi ^T}(t)]^T}$, it can be obtained from (\ref{8}) and (\ref{9}) that
\begin{eqnarray}\label{10}
\xi (t) = {\xi _a}(t) + {\xi _d}(t).
\end{eqnarray}
Since $U \otimes {I_{2n}}$ is nonsingular, one can concluded that ${\xi _a}(t)$ and ${\xi _d}(t)$ are linearly independent. It follows from (\ref{8}) and the fact ${[{\kappa ^T}(t),0]^T} = {e_1} \otimes \kappa (t)$ that
\begin{eqnarray}\label{11}
{\xi _a}(t) = \left( {U \otimes {I_{2n}}} \right)\left( {{e_1} \otimes \kappa (t)} \right) = U{e_1} \otimes \kappa (t) = {\mathbf{1}_N} \otimes \kappa (t).
\end{eqnarray}
From (\ref{10}) and (\ref{11}), one can show that
\begin{eqnarray}\label{12}
{\xi _d}(t) = \xi (t) - {\mathbf{1}_N} \otimes \kappa (t).
\end{eqnarray}
From (\ref{9}), (\ref{10}) and (\ref{12}), one can find that for any given positive constant $\varepsilon $, there exists a finite constant ${t_{\varepsilon }}$ such that $\left\| {{x_i}(t) - {f_i}(t) - \kappa (t)} \right\| \leqslant \varepsilon $ $(i \in \{ 1,2, \cdots ,N\} )$, $\forall t \geqslant {t_\varepsilon}$, if $\left\| {\varphi (t)} \right\| \leqslant {\varepsilon  \mathord{\left/ {\vphantom {\varepsilon  {{\varepsilon _N}}}} \right. \kern-\nulldelimiterspace} {{\varepsilon _N}}} = {\varepsilon _\varphi }$, $\forall t \geqslant {t_\varepsilon}$, which means that $\varphi (t)$ represents the time-varying formation error and $\kappa (t)$ shows one of the candidates of formation center functions, respectively.\par

From (\ref{8}), one can obtain that \vspace{-5pt}
\begin{eqnarray}\label{13}
\kappa (0) = ({\bar u_1} \otimes {I_{2n}})(x(0) - f(0)).
\end{eqnarray}
One can show that
\[\begin{gathered}
  \int_0^t {{e^{({\theta _1}\theta _2^T + {\theta _2}\alpha )(t - \tau )}}\left( {{{\bar u}_1} \otimes {\theta _1}} \right)\left( {{f_v}(\tau ) - {{\dot f}_p}(\tau )} \right)} d\tau  \hfill \\
   = \int_0^t {{e^{({\theta _1}\theta _2^T + {\theta _2}\alpha )(t - \tau )}}\left( {{{\bar u}_1} \otimes {\theta _1}} \right){f_v}(\tau )} d\tau  + {e^{({\theta _1}\theta _2^T + {\theta _2}\alpha )t}}\left( {{{\bar u}_1} \otimes {\theta _1}} \right){f_p}(0) \hfill \\
\end{gathered} \]
\vspace{-2em}
\begin{eqnarray}\label{14}
  \hspace{5em} - \int_0^t {{e^{({\theta _1}\theta _2^T + {\theta _2}\alpha )(t - \tau )}}\left( {{{\bar u}_1} \otimes ({\theta _1}\theta _2^T + {\theta _2}\alpha ){\theta _1}} \right){f_p}(\tau )} d\tau  - \left( {{{\bar u}_1} \otimes {\theta _1}} \right){f_p}(t),
\end{eqnarray}
and
\[\begin{gathered}
  \int_0^t {{e^{({\theta _1}\theta _2^T + {\theta _2}\alpha )(t - \tau )}}\left( {{{\bar u}_1} \otimes {\theta _2}} \right){{\dot f}_v}(\tau )} d\tau  \hfill \\
   = \left( {{{\bar u}_1} \otimes {\theta _2}} \right){f_v}(t) - {e^{({\theta _1}\theta _2^T + {\theta _2}\alpha )t}}\left( {{{\bar u}_1} \otimes {\theta _2}} \right){f_v}(0) \hfill \\
\end{gathered} \]
\vspace{-2em}
\begin{eqnarray}\label{15}
  \hspace{7em} + \int_0^t {{e^{({\theta _1}\theta _2^T + {\theta _2}\alpha )(t - \tau )}}\left( {{{\bar u}_1} \otimes ({\theta _1}\theta _2^T + {\theta _2}\alpha ){\theta _2}} \right){f_v}(\tau )} d\tau .
\end{eqnarray}
By the structure of $f(t)$, it can be found that
\begin{eqnarray}\label{16}
\left( {{{\bar u}_1} \otimes {\theta _1}} \right){f_p}(t) + \left( {{{\bar u}_1} \otimes {\theta _2}} \right){f_v}(t) = \left( {{{\bar u}_1} \otimes {I_{2n}}} \right)f(t),
\end{eqnarray}
\begin{eqnarray}\label{17}
{e^{({\theta _1}\theta _2^T + {\theta _2}\alpha )t}}\left( {\left( {{{\bar u}_1} \otimes {\theta _1}} \right){f_p}(0) + \left( {{{\bar u}_1} \otimes {\theta _2}} \right){f_v}(0)} \right) = {e^{({\theta _1}\theta _2^T + {\theta _2}\alpha )t}}\left( {{{\bar u}_1} \otimes {I_{2n}}} \right)f(0).
\end{eqnarray}
Then, it follows from (\ref{14})-(\ref{17}) that
\[\begin{gathered}
  \int_0^t {{e^{({\theta _1}\theta _2^T + {\theta _2}\alpha )(t - \tau )}}\left( {{{\bar u}_1} \otimes {\theta _1}} \right)\left( {{f_v}(\tau ) - {{\dot f}_p}(\tau )} \right)} d\tau  \hfill \\
   = \int_0^t {{e^{({\theta _1}\theta _2^T + {\theta _2}\alpha )(t - \tau )}}\left( {{{\bar u}_1} \otimes {\theta _2}} \right)\left( {{{\dot f}_v}(\tau ) - {\alpha _p}{f_p}(\tau ) - {\alpha _v}{f_v}(\tau )} \right)} d\tau  \hfill \\
\end{gathered} \]
\vspace{-2em}
\begin{eqnarray}\label{18}
  \hspace{-2em} + {e^{({\theta _1}\theta _2^T + {\theta _2}\alpha )t}}\left( {{{\bar u}_1} \otimes {I_{2n}}} \right)f(0) - \left( {{{\bar u}_1} \otimes {I_{2n}}} \right)f(t).
\end{eqnarray}
In virtue of (\ref{6}), (\ref{13}) and (\ref{18}), the conclusion of Theorem \ref{theorem1} can be obtained.
\end{proof}

\section{Robust time-varying formation design }\label{section4}\vspace{-0.5em}

~~~~In this section, firstly, an algorithm is presented to show the procedure of designing the robust time-varying formation control protocol. Then, sufficient conditions of the robust time-varying formation design are shown and the tracking performance and the robustness stability of multi-agent systems are analyzed, respectively.\par

The core idea of designing robust time-varying formation control protocol (\ref{2}) is to determine the gain matrix and the robust disturbance compensation. The following algorithm with four steps is presented to design protocol (\ref{2}).\vspace{10pt}

\hspace{-10pt}{\it Robust Time-Varying Formation Design Algorithm}\par

Step 1: Check the following formation feasibility condition for the expected time-varying formation.
\vspace{-10pt}
\begin{eqnarray}\label{19}
{\left\| {{f_{iv}}(t) - {{\dot f}_{ip}}(t)} \right\|_\infty } \leqslant {\varepsilon _f},{\text{ }}\forall t \geqslant {t_\varepsilon },{\text{ }}\forall i \in \{ 1,2, \cdots ,N\} .
\end{eqnarray}
If condition (\ref{19}) is satisfied, then go to Step 2; else the expected time-varying formation cannot be reached by multi-agent system (\ref{1}) with protocol (\ref{2}) and the algorithm stops.\par
Step 2: Solve the following algebraic Riccati equation for a positive definite matrix $P$
\vspace{-10pt}
\begin{eqnarray}\label{20}
P({\theta _1}\theta _2^T + {\theta _2}\alpha ) + {({\theta _1}\theta _2^T + {\theta _2}\alpha )^T}P - P{\theta _2}\theta _2^TP + I = 0.
\end{eqnarray}\par
Step 3: Set the gain matrix $K_u$ as ${K_u} = {\operatorname{Re} ^{ - 1}}({\lambda _2})\theta _2^TP$.\par
Step 4: Choose sufficiently large bandwidth constants ${\beta _{ig}}$ and ${\beta _{iz}}$ $(i \in \{ 1,2, \cdots ,N\} )$ of ESO (\ref{3}) to effectively estimate the robust disturbance compensation.\par

With the robust time-varying formation design algorithm, tracking performances and robustness stability properties are analyzed in the following theorem.\par

\begin{thm} \label{theorem2}
For any given bounded initial states, if formation feasibility condition (\ref{19}) is satisfied, then multi-agent system (\ref{1}) reaches the robust time-varying formation by protocol (\ref{2}) designed in the robust time-varying formation design algorithm.
\end{thm}\vspace{-10pt}

\begin{proof}
Firstly, consider the stability of the following subsystem:
\begin{eqnarray}\label{21}
{\dot \eta _k}(t) = \left( {{\theta _1}\theta _2^T + {\theta _2}\alpha  - {\lambda _k}{\theta _2}{K_u}} \right){\eta _k}(t),
\end{eqnarray}
where $\forall k \in \{ 2,3, \cdots ,N\} $. Then construct the Lyapunov function candidate as follows:
\begin{eqnarray}\label{22}
{V_k}(t) = \eta _k^H(t)P{\eta _k}(t).
\end{eqnarray}
Let ${K_u} = {\operatorname{Re} ^{ - 1}}({\lambda _2})\theta _2^TP$, then differentiating $V(t)$ along the trajectories of (\ref{21}) yields
\begin{eqnarray}\label{23}
{\dot V_k}(t) = \eta _k^H(t)\left( {{{({\theta _1}\theta _2^T + {\theta _2}\alpha )}^T}P + P({\theta _1}\theta _2^T + {\theta _2}\alpha ) - 2\operatorname{Re} ({\lambda _k}){{\operatorname{Re} }^{ - 1}}({\lambda _2})P{\theta _2}\theta _2^TP} \right){\eta _k}(t).
\end{eqnarray}
Substituting $P({\theta _1}\theta _2^T + {\theta _2}\alpha ) + {({\theta _1}\theta _2^T + {\theta _2}\alpha )^T}P = P{\theta _2}\theta _2^TP - I$ into (\ref{23}) gives
\vspace{-5pt}
\begin{eqnarray}\label{24}
{\dot V_k}(t) = \eta _k^H(t)\left( {\left( {1 - 2\operatorname{Re} ({\lambda _k}){{\operatorname{Re} }^{ - 1}}({\lambda _2})} \right)P{\theta _2}\theta _2^TP - I} \right){\eta _k}(t).
\end{eqnarray}
Due to $0 < \operatorname{Re} ({\lambda _2}) \leqslant  \cdots  \leqslant \operatorname{Re} ({\lambda _N})$, one can derive from (\ref{24}) that ${\dot V_k}(t) \leqslant  - \eta _k^H(t){\eta _k}(t)$ ($\forall k \in \{ 2,3, \cdots ,N\} $). Therefore, ${\eta _k}(t)$ converges to $0$ asymptotically, which means that ${\theta _1}\theta _2^T + {\theta _2}\alpha  - {\lambda _k}{\theta _2}{K_u}$ is Hurwitz. By the structure of $\tilde J$, one can conclude that ${I_{N - 1}} \otimes ({\theta _1}\theta _2^T + {\theta _2}\alpha ) - \tilde J \otimes {\theta _2}{K_u}$ is Hurwitz.\par

Then, the tracking performance and the robustness stability is analyzed. Let $A = {I_{N - 1}} \otimes ({\theta _1}\theta _2^T + {\theta _2}\alpha ) - \tilde J \otimes {\theta _2}{K_u}$, then subsystem (\ref{7}) can be rewritten as
\begin{eqnarray}\label{25}
\dot \varphi (t) = A\varphi (t) + (\bar u \otimes {\theta _2})\left( {\omega (t) - z(t)} \right) + \left( {\bar u \otimes {\theta _1}} \right)\left( {{f_v}(t) - {{\dot f}_p}(t)} \right).
\end{eqnarray}
By Laplace transform, (\ref{3}) can be converted to
\begin{eqnarray}\label{26}
\left\{ \begin{gathered}
  {z_i}(s) - {\omega _i}(s) + ({\beta _{ig}} + s)\left( {{v_i}(s) - {g_i}(s)} \right) = 0, \hfill \\
  s{z_i}(s) + {\beta _{iz}}{g_i}(s) - {\beta _{iz}}{v_i}(s) = 0, \hfill \\
\end{gathered}  \right.
\end{eqnarray}
where $i \in \{ 1,2, \cdots ,N\} $. From (\ref{26}), it can be shown that
\begin{eqnarray}\label{27}
{z_i}(s) = {G_i}(s){\omega _i}(s),
\end{eqnarray}
where ${G_i}(s) = {{{\beta _{iz}}} \mathord{\left/ {\vphantom {{{\beta _{iz}}} {({s^2} + {\beta _{ig}}s + {\beta _{iz}})}}} \right. \kern-\nulldelimiterspace} {({s^2} + {\beta _{ig}}s + {\beta _{iz}})}}$, $i \in \{ 1,2, \cdots ,N\} $. Let ${\beta _{ig}} = 2{\sigma _i}$ and ${\beta _{iz}} = \sigma _i^2$, then one can obtain that
\begin{eqnarray}\label{28}
{G_i}(s) = \frac{{\sigma _i^2}}{{{{(s + {\sigma _i})}^2}}}.
\end{eqnarray}
From (\ref{27}), it follows that
\begin{eqnarray}\label{29}
  \omega (s) - z(s) = {\text{diag}}\{ 1 - {G_1}(s),1 - {G_2}(s), \cdots ,1 - {G_N}(s)\} \omega (s) = {\Phi _N}(s)\omega (s).
\end{eqnarray}
Define
\begin{eqnarray}\label{30}
\left\{ \begin{gathered}
  {\rho _\omega } = {\left\| {{{(s{I_{2n(N - 1)}} - A)}^{ - 1}}\left( {\bar u{\Phi _N}(s) \otimes {\theta _2}} \right)} \right\|_1}, \hfill \\
  {\rho _f} = {\left\| {{{(s{I_{2n(N - 1)}} - A)}^{ - 1}}\left( {\bar u \otimes {\theta _1}} \right)} \right\|_1}, \hfill \\
  {\upsilon _{ \varphi (0)}} = {\left\| {{e^{At}}\varphi (0)} \right\|_\infty }. \hfill \\
\end{gathered}  \right.
\end{eqnarray}
From (\ref{25}), (\ref{27}), (\ref{29}) and (\ref{30}), it can be derived that
\begin{eqnarray}\label{31}
{\left\| {\varphi (t)} \right\|_\infty } \leqslant {\upsilon _{\varphi (0)}} + {\rho _\omega }{\left\| {\omega (t)} \right\|_\infty } + {\rho _f}{\left\| {{f_v}(t) - {{\dot f}_p}(t)} \right\|_\infty }.
\end{eqnarray}
For agent $i$ $(i \in \{ 1,2, \cdots ,N\} )$, since $\omega (t)$ is bounded, there exist two positive constants ${\gamma _{ \varphi i}}$ and ${\delta _{\omega \varphi i}}$ such that
\begin{eqnarray}\label{32}
{\left\| {{\omega _i}(t)} \right\|_\infty } \leqslant {\gamma _{ \varphi i}}{\left\| {{u_i}(t)} \right\|_\infty } + {\delta _{\omega \varphi i}},{\text{ }}i \in \{ 1,2, \cdots ,N\} .
\end{eqnarray}
It follows from (\ref{32}) that positive constants ${\gamma _{ \varphi }}$ and ${\delta _{\omega  \varphi }}$ exist such that
\begin{eqnarray}\label{33}
{\left\| {\omega (t)} \right\|_\infty } \leqslant {\gamma _{ \varphi}}{\left\| {u(t)} \right\|_\infty } + {\delta _{\omega  \varphi }}.
\end{eqnarray}
By (\ref{2}), (\ref{3}) and (\ref{29}), one can show that
\begin{eqnarray}\label{34}
{\left\| {u(t)} \right\|_\infty } = {\delta _{u \varphi 1}}{\left\| {\varphi (t)} \right\|_\infty } + {\delta _{u \varphi 2}}{\left\| {\omega (t)} \right\|_\infty } + {\delta _{u \varphi 3}},
\end{eqnarray}
where ${\delta _{u \varphi 1}}$, ${\delta _{u \varphi 2}}$ and ${\delta _{u \varphi 3}}$ are positive constants. Substituting (\ref{34}) into (\ref{33}), one can obtain that ${\upsilon _{\varphi }}$ and ${\upsilon _e}$ exist such that
\begin{eqnarray}\label{35}
{\left\| {\omega (t)} \right\|_\infty } \leqslant {\upsilon _{\varphi }}{\left\| {\varphi (t)} \right\|_\infty } + {\upsilon _e}.
\end{eqnarray}
If ${\left\| {\bar u \otimes {\theta _2}} \right\|_\infty }$ is bounded and ${\sigma _i}$ $(i = 1,2, \cdots ,N)$ are sufficiently large, then it can be deduced from (\ref{31}) and (\ref{35}) that
\begin{eqnarray}\label{36}
\left\{ \begin{gathered}
  {\left\| {\omega (t)} \right\|_\infty } \leqslant \frac{{{\upsilon _{\varphi }}{\upsilon _{\varphi (0)}} + {\upsilon _e}}}
{{1 - {\upsilon _{\varphi }}{\rho _\omega }}}, \hfill \\
  {\left\| {\varphi (t)} \right\|_\infty } \leqslant \frac{{{\upsilon _{ \varphi (0)}} + {\upsilon _e}{\rho _\omega }}}
{{1 - {\upsilon _{ \varphi }}{\rho _\omega }}}. \hfill \\
\end{gathered}  \right.
\end{eqnarray}
It follows from (\ref{36}) that
\begin{eqnarray}\label{37}
\left\{ \begin{gathered}
  {\left\| {\omega (t)} \right\|_\infty } \leqslant {{\tilde \upsilon }_\omega }, \hfill \\
  {\left\| {\varphi (t)} \right\|_\infty } \leqslant {{\tilde \upsilon }_{\varphi }}, \hfill \\
\end{gathered}  \right.
\end{eqnarray}
where ${\tilde \upsilon _\omega }$ and ${\tilde \upsilon _{ \varphi }}$ are positive constants. According to the formation feasibility condition (\ref{19}), one has that
\begin{eqnarray}\label{38}
{\left\| {{f_v}(t) - {{\dot f}_p}(t)} \right\|_\infty } \leqslant {\varepsilon _f},{\text{ }}\forall t \geqslant {t_f}.
\end{eqnarray}
From (\ref{31}), (\ref{37}) and (\ref{38}), one can obtain that
\begin{eqnarray}\label{39}
\mathop {\max }\limits_i \left| {{\varphi _i}(t)} \right| \leqslant \mathop {\max }\limits_i \left| {c_{_{2n(N - 1),i}}^T{e^{At}}\varphi (0)} \right| + {\rho _\omega }{\tilde \upsilon _\omega } + {\rho _f}{\varepsilon _f},{\text{ }}\forall t \geqslant {t_f},
\end{eqnarray}
where $i \in \{ 2,3, \cdots ,N\} $, ${c_{2n(N - 1),i}}$ is a $2n(N - 1)$-dimensional unit column vector with the $i$th element $1$ and other elements $0$. For the bounded initial states ${ \varphi _i}(0)$ $(i \in \{ 2,3, \cdots ,N\} )$, one can find that ${\varphi _i}(t)$ is bounded. It can also be obtained that the states of the robust disturbance compensation ${z_i}(t)$ and the control protocol ${u_i}(t)$ are bounded. It follows that all states involved in the closed-loop system (\ref{4}) are bounded. Furthermore, since $A$ is Hurwitz, there exists a finite constant ${t_\varepsilon } \geqslant {t_f}$ such that $\left\| {\varphi (t)} \right\| \leqslant {\varepsilon _\varphi }$, $\forall t \geqslant {t_\varepsilon }$ for any given positive constant ${\varepsilon _\varphi }$, which means that multi-agent system (\ref{1}) is robust time-varying formation-reachable by protocol (\ref{2}). This completes the proof.
\end{proof}
\vspace{-10pt}

\section{Numerical simulations}\label{section5}
\vspace{-5pt}
~~~~In this section, a simulation example is provided to demonstrate the effectiveness of the theoretical results obtained in previous sections. \par

Consider a second-order multi-agent system containing six agents in the $XYZ$ space ($n = 3$), where the interaction topology among agents is described as a 0-1 weighted digraph in Figure 1. The dynamics of each agent can be described by (\ref{1}) with ${\alpha _p} =  - 0.01$ and ${\alpha _v} = 0$. Let ${x_i}(t) = {\left[ {{p_{iX}}(t),{p_{iY}}(t),{p_{iZ}}(t),{v_{iX}}(t),{v_{iY}}(t),{v_{iZ}}(t)} \right]^T}$ $(i \in \{ 1,2, \cdots ,6\} )$, where ${p_{iX}}(t)$, ${p_{iY}}(t)$, ${p_{iZ}}(t)$ and ${v_{iX}}(t)$, ${v_{iY}}(t)$, ${v_{iZ}}(t)$ are positions and velocities along the $X$ axis, $Y$ axis and $Z$ axis, respectively. The initial states of each agent are set as ${x_1}(t) = {\left[ {0.6,1.2,0.5, - 1.2, - 0.3,0.8} \right]^T}$, ${x_2}(t) = {\left[ { - 1.5, - 0.3,1.8, - 1.6,2.3,1.1} \right]^T}$, ${x_3}(t) = {\left[ {2.1,0.8, - 1.6,0.3, - 1.9,2.5} \right]^T}$, ${x_4}(t) = {\left[ {3.8,1.7, - 2.6,1.8, - 3.3,1.5} \right]^T}$, ${x_5}(t) = {\left[ {4.5,1.9, - 1.2, - 2.9,3.5, - 1.4} \right]^T}$ and ${x_6}(t) = {\left[ { - 4.2,2.9,3.8, - 5.1, - 3.5,2.7} \right]^T}$.\par

\begin{figure}[!htb]
\begin{center}
\scalebox{1.1}[1.1]{\includegraphics{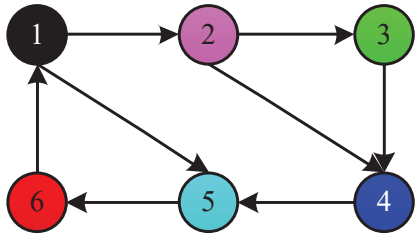}}
\vspace{0em}
\caption{Interaction topology $G$.}
\end{center}\vspace{0em}
\end{figure}
The six agents are required to follow a time-varying formation in the form of
\[{f_i}(t) = 3\left[ \begin{gathered}
  \sin (t + {{(i - 1)\pi } \mathord{\left/
 {\vphantom {{(i - 1)\pi } 3}} \right.
 \kern-\nulldelimiterspace} 3}) \hfill \\
  \cos (t + {{(i - 1)\pi } \mathord{\left/
 {\vphantom {{(i - 1)\pi } 3}} \right.
 \kern-\nulldelimiterspace} 3}) \hfill \\
   - \sin (t + {{(i - 1)\pi } \mathord{\left/
 {\vphantom {{(i - 1)\pi } 3}} \right.
 \kern-\nulldelimiterspace} 3}) \hfill \\
  \cos (t + {{(i - 1)\pi } \mathord{\left/
 {\vphantom {{(i - 1)\pi } 3}} \right.
 \kern-\nulldelimiterspace} 3}) \hfill \\
   - \sin (t + {{(i - 1)\pi } \mathord{\left/
 {\vphantom {{(i - 1)\pi } 3}} \right.
 \kern-\nulldelimiterspace} 3}) \hfill \\
   - \cos (t + {{(i - 1)\pi } \mathord{\left/
 {\vphantom {{(i - 1)\pi } 3}} \right.
 \kern-\nulldelimiterspace} 3}) \hfill \\
\end{gathered}  \right],{\text{ }}(i = 1,2, \cdots ,6).\]
It can be found from ${f_i}(t)$ that both positions and velocities of six agents can take shape regular hexagons with time-varying edges. One can see that formation feasibility condition (\ref{19}) is satisfied. The external disturbances are generated by

\[{\omega _i}(t) = \left[ \begin{gathered}
  (2.5 + 0.2(i - 1))\sin t + 1.5 + 1.2(i - 1) \hfill \\
  (1.5 + 0.2(i - 1))\sin t + 2.5 + 1.2(i - 1) \hfill \\
  (2 + 0.2(i - 1))\sin (t + 0.4\pi ) + 3 + 0.2(i - 1) \hfill \\
\end{gathered}  \right],{\text{ }}(i = 1,2, \cdots ,6).\]

Choose the bandwidth constants of ESO (\ref{3}) as ${\beta _{ig}} = 2{\sigma _i}$ and ${\beta _{iz}} = \sigma _i^2$ with ${\sigma _i} = 10$ $(i \in \{ 1,2, \cdots ,6\} )$. From Theorem \ref{theorem2}, one can determine that ${K_u} = [1.0654,1.8576] \otimes {I_3}$.\par

Figures 2 and 3 describe the position and velocity trajectory of six agents and the formation center at $t = 0$s, $t = 10$s, $t = 15$s and $t = 20$s, respectively, where the position and velocity states of agents are represented by asterisks, plus signs, circles, x marks, pentagrams and squares, and the formation centers are denoted by hexagrams. Figure 4 presents the curves of the formation centers for positions and velocities within $t = 20$s, where the initial and final states are depicted by circles and squares, respectively. Figure 5 shows the trajectory of the formation center within $t = 20$s.\par

From Figures 2(a)-(b) and 3(a)-(b), one can find that the multi-agent system can achieve the regular pentagon formation in both position and velocity states. Figures 2(b)-(d) and 3(b)-(d) present that the formation keeps rotation in position and velocity states, respectively; that is, the formation is time-varying. From the simulation results shown in Figures 2-5, one can conclude that second-order multi-agent system (\ref{1}) with external disturbance achieves the robust time-varying formation by protocol (\ref{2}).\par

\begin{figure}[!htb]
\begin{center}
\scalebox{0.4}[0.4]{\includegraphics{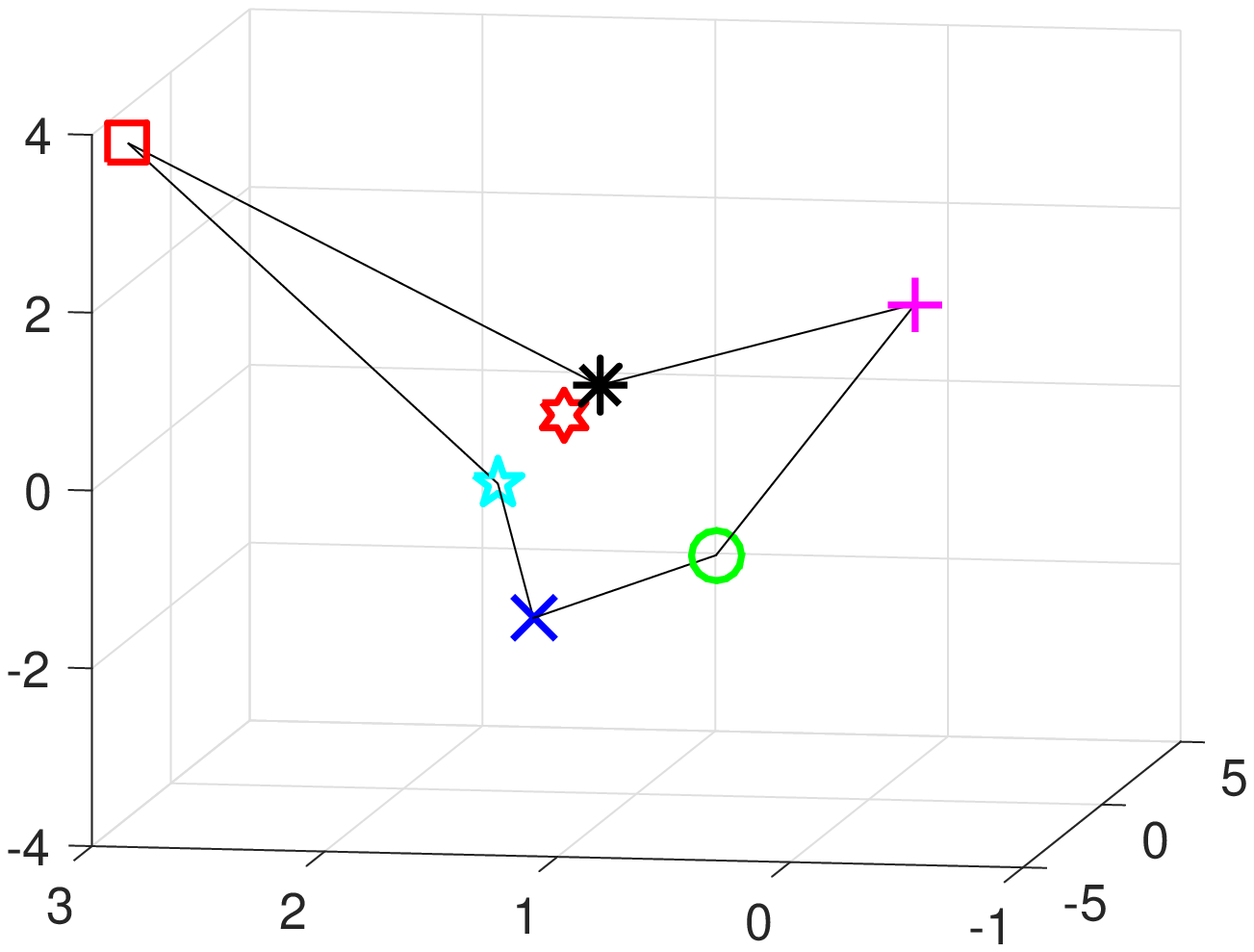}}
\scalebox{0.4}[0.4]{\includegraphics{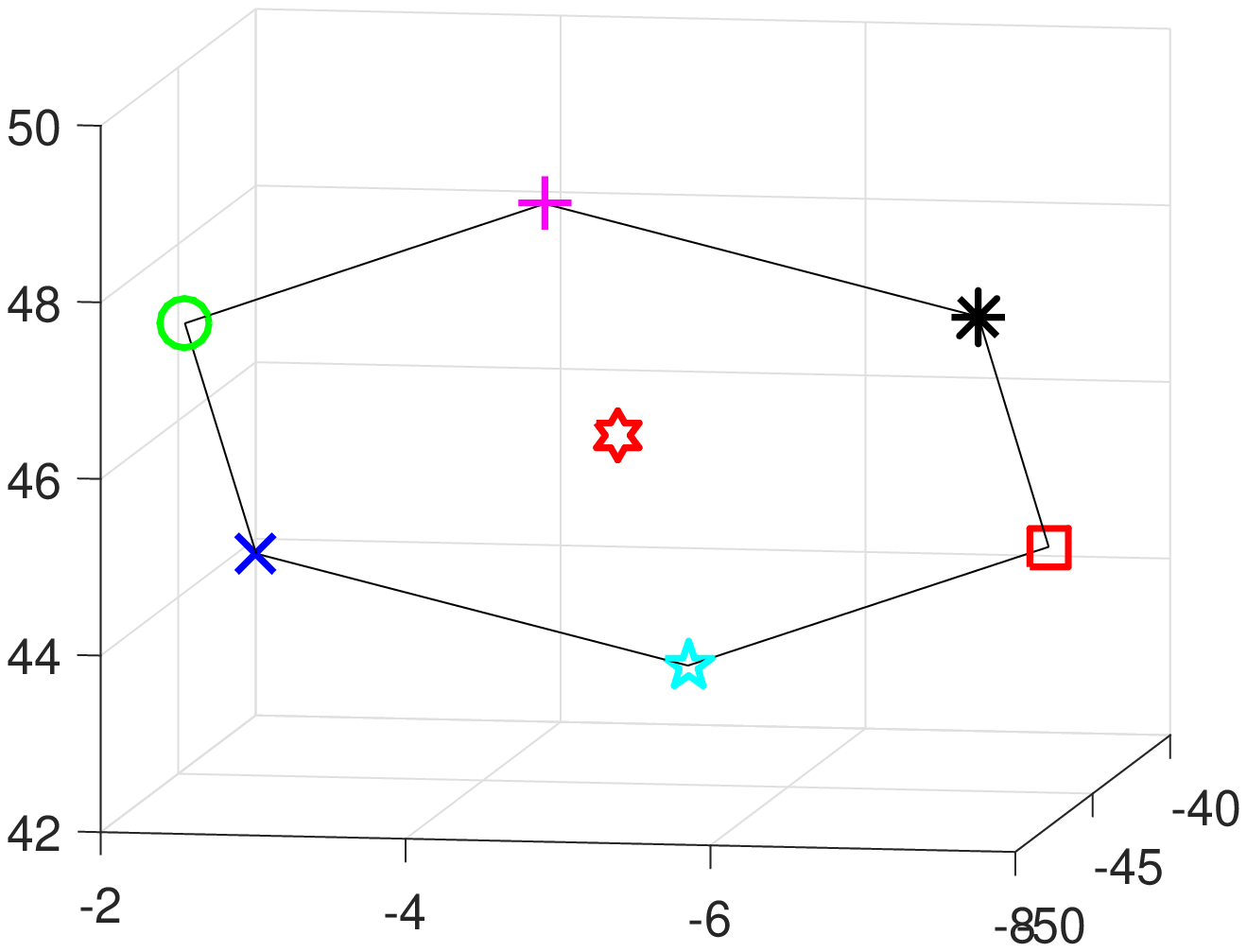}}
\put (-340, 50){\rotatebox{90} {{\scriptsize ${p_{iZ}}(t)$}}}
\put (-273, 0){ {{\scriptsize ${p_{iY}}(t)$}}}
\put (-193, 10){ {{\scriptsize ${p_{iX}}(t)$}}}
\put (-276, -12) {{ \scriptsize (a)~{\it t} = 0s}}
\put (-173, 50){\rotatebox{90} {{\scriptsize ${p_{iZ}}(t)$}}}
\put (-106, 0){ {{\scriptsize ${p_{iY}}(t)$}}}
\put (-16, 10){ {{\scriptsize ${p_{iX}}(t)$}}}
\put (-109, -12) {{ \scriptsize (b)~{\it t} = 10s}}
\end{center}\vspace{-2em}
\end{figure}
\begin{figure}[!htb]
\begin{center}
\scalebox{0.4}[0.4]{\includegraphics{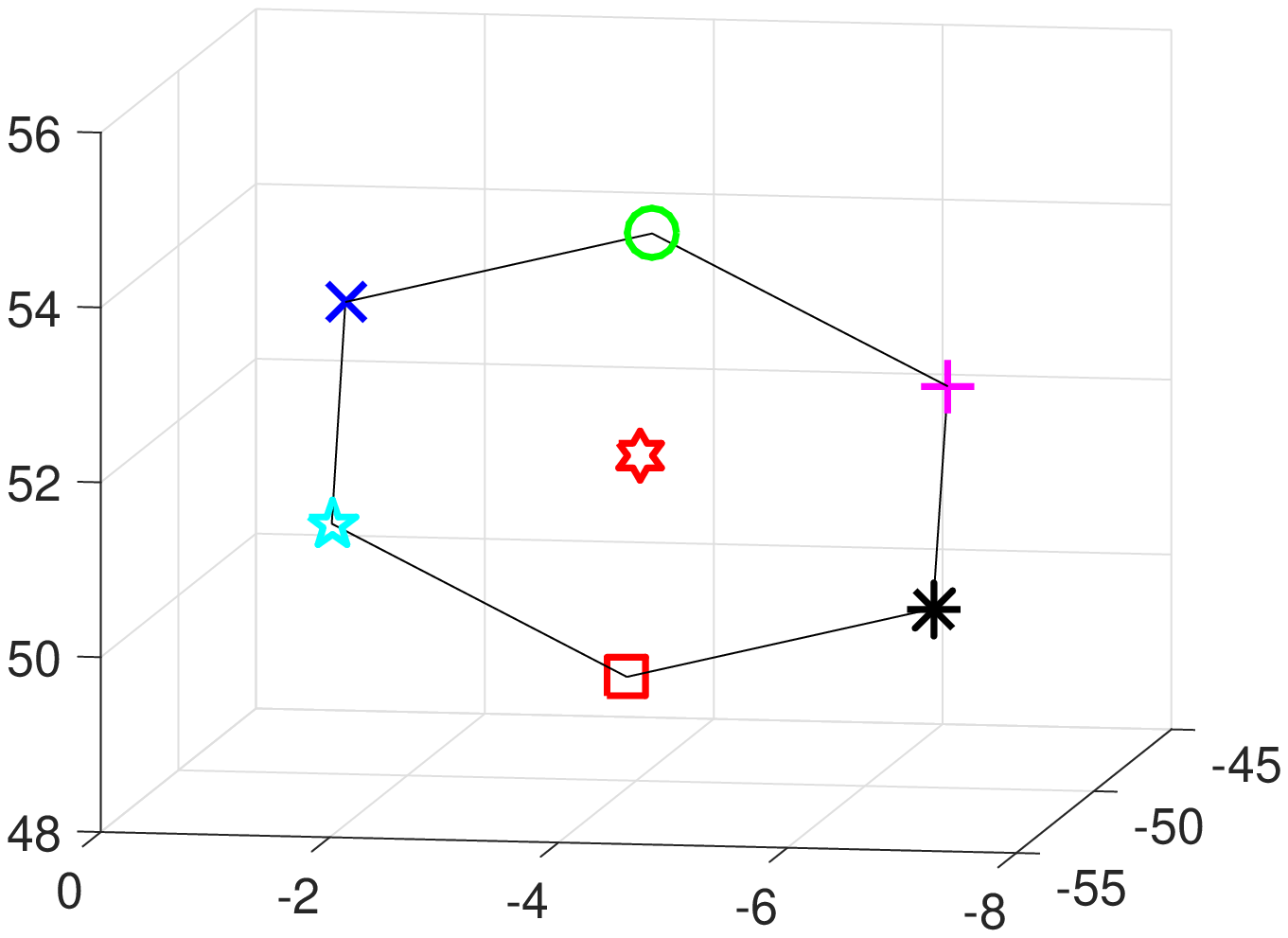}}
\scalebox{0.4}[0.4]{\includegraphics{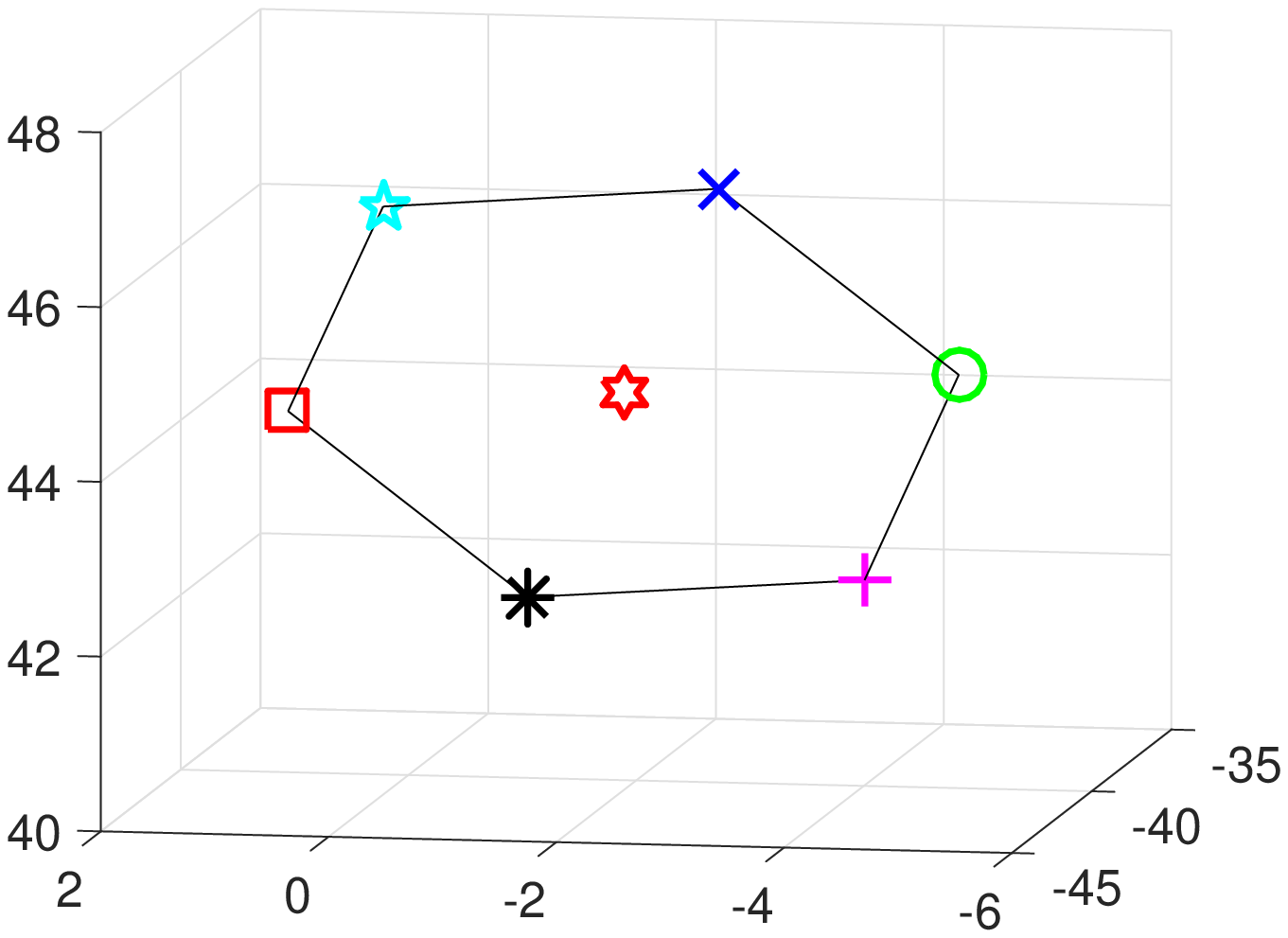}}
\put (-340, 50){\rotatebox{90} {{\scriptsize ${p_{iZ}}(t)$}}}
\put (-273, 0){ {{\scriptsize ${p_{iY}}(t)$}}}
\put (-193, 10){ {{\scriptsize ${p_{iX}}(t)$}}}
\put (-276, -12) {{ \scriptsize (c)~{\it t} = 15s}}
\put (-173, 50){\rotatebox{90} {{\scriptsize ${p_{iZ}}(t)$}}}
\put (-106, 0){ {{\scriptsize ${p_{iY}}(t)$}}}
\put (-16, 10){ {{\scriptsize ${p_{iX}}(t)$}}}
\put (-109, -12) {{ \scriptsize (d)~{\it t} = 20s}}
\vspace{0em} \caption{Position curves of six agents and the formation center at different time.}
\end{center}\vspace{-2em}
\end{figure}

\begin{figure}[!htb]
\begin{center}
\scalebox{0.4}[0.4]{\includegraphics{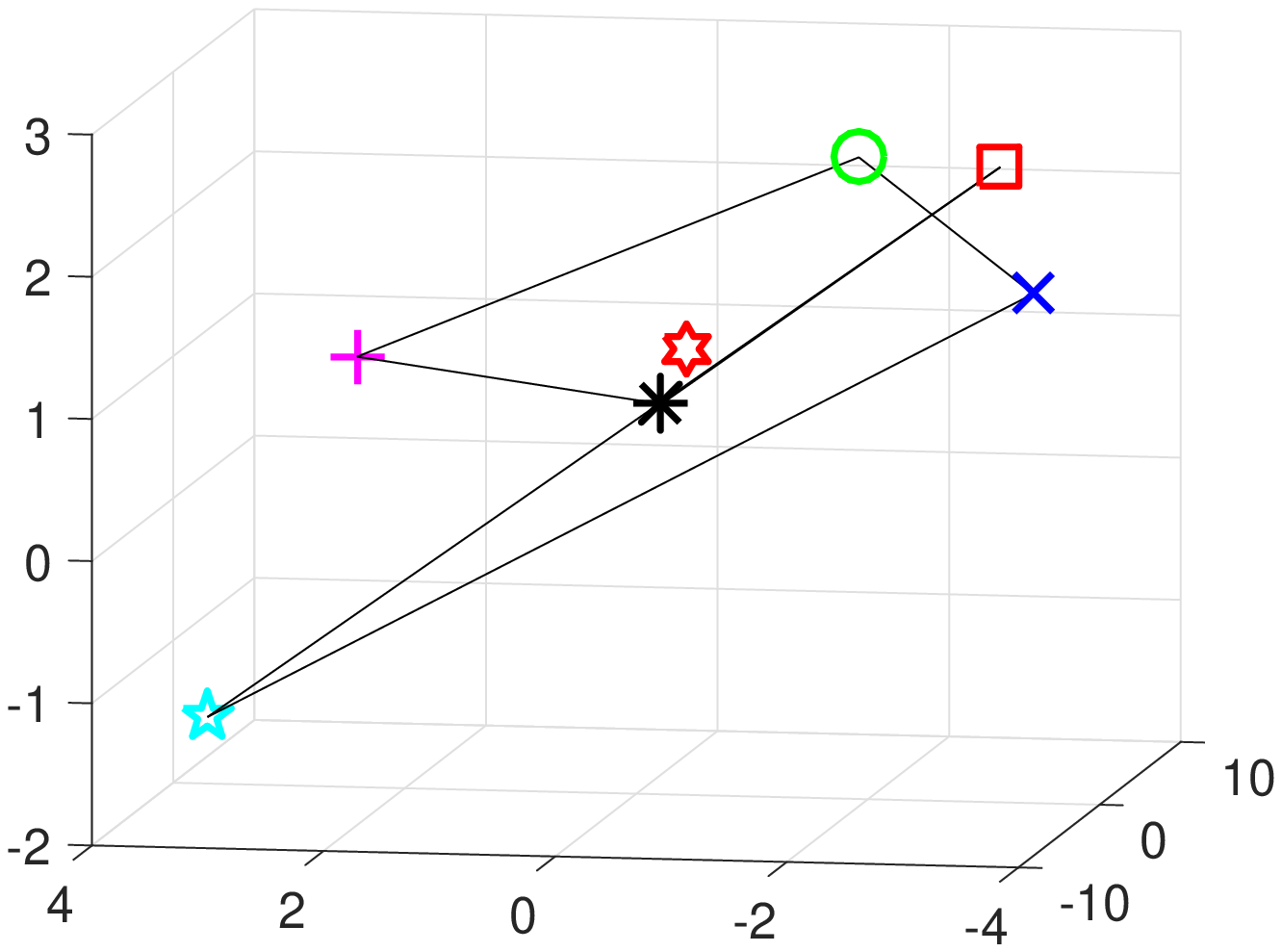}}
\scalebox{0.4}[0.4]{\includegraphics{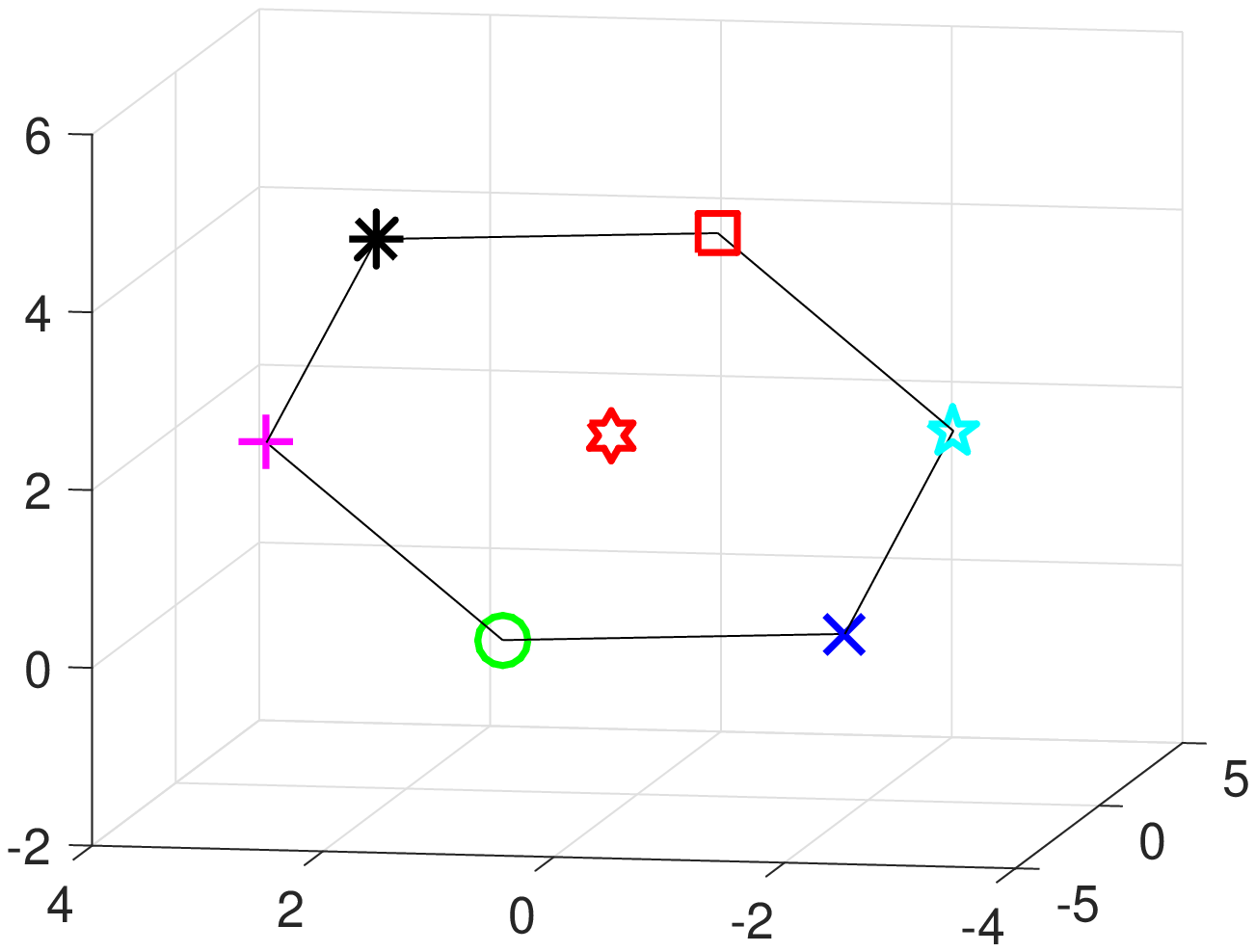}}
\put (-340, 50){\rotatebox{90} {{\scriptsize ${v_{iZ}}(t)$}}}
\put (-273, 0){ {{\scriptsize ${v_{iY}}(t)$}}}
\put (-193, 10){ {{\scriptsize ${v_{iX}}(t)$}}}
\put (-276, -12) {{ \scriptsize (a)~{\it t} = 0s}}
\put (-173, 50){\rotatebox{90} {{\scriptsize ${v_{iZ}}(t)$}}}
\put (-106, 0){ {{\scriptsize ${v_{iY}}(t)$}}}
\put (-16, 10){ {{\scriptsize ${v_{iX}}(t)$}}}
\put (-109, -12) {{ \scriptsize (b)~{\it t} = 10s}}
\end{center}\vspace{-2em}
\end{figure}
\begin{figure}[!htb]
\begin{center}
\scalebox{0.4}[0.4]{\includegraphics{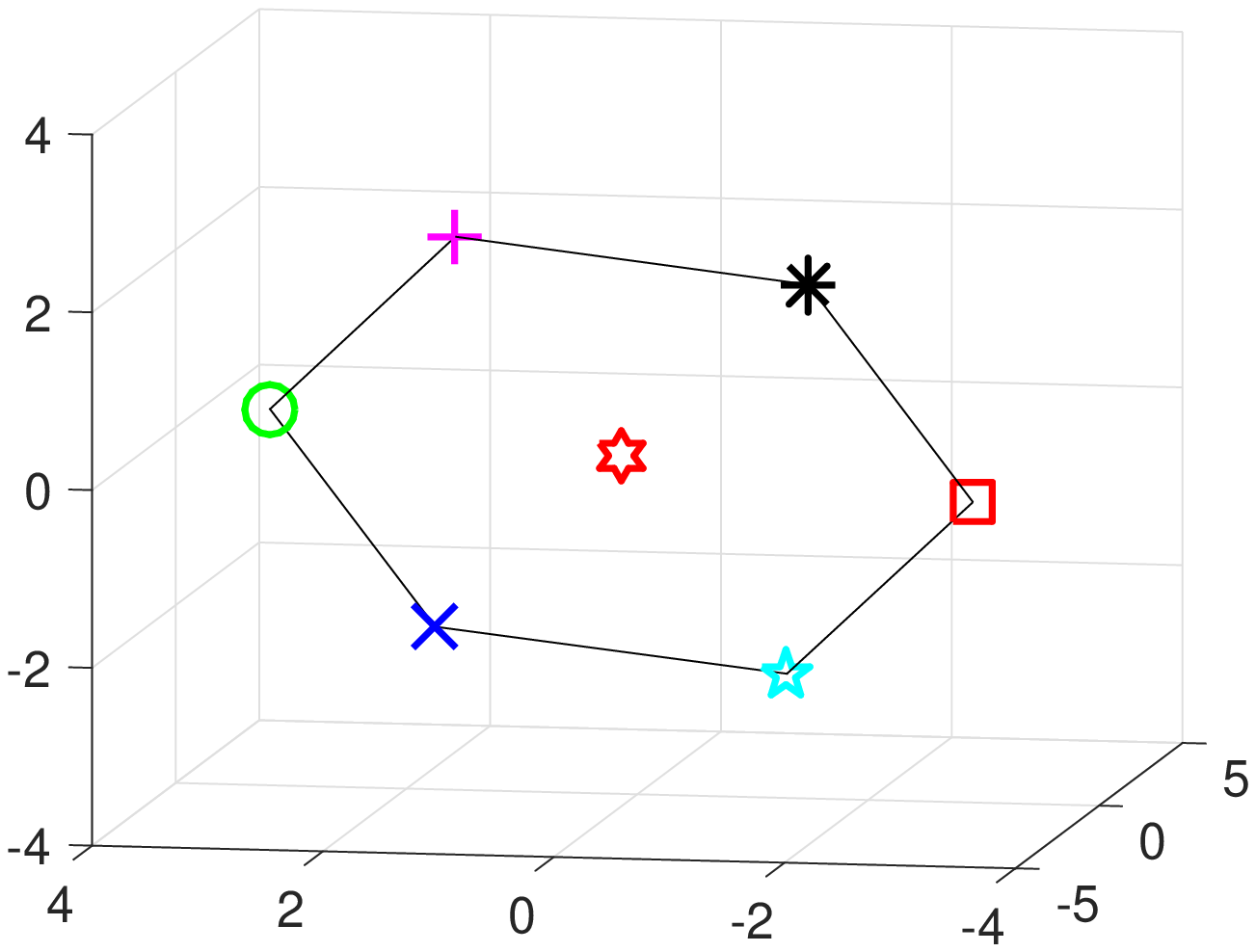}}
\scalebox{0.4}[0.4]{\includegraphics{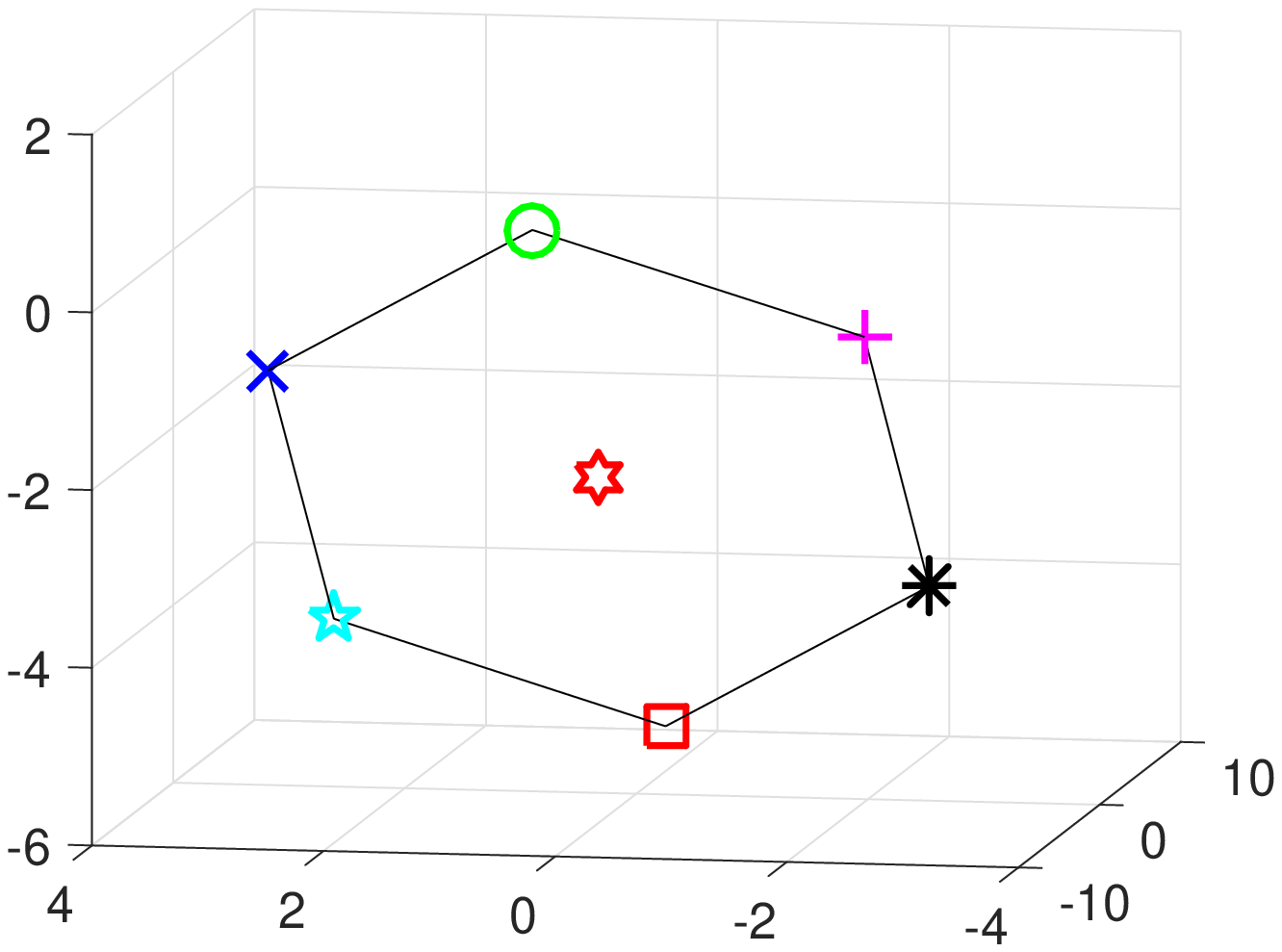}}
\put (-340, 50){\rotatebox{90} {{\scriptsize ${v_{iZ}}(t)$}}}
\put (-273, 0){ {{\scriptsize ${v_{iY}}(t)$}}}
\put (-193, 10){ {{\scriptsize ${v_{iX}}(t)$}}}
\put (-276, -12) {{ \scriptsize (c)~{\it t} = 15s}}
\put (-173, 50){\rotatebox{90} {{\scriptsize ${v_{iZ}}(t)$}}}
\put (-106, 0){ {{\scriptsize ${v_{iY}}(t)$}}}
\put (-16, 10){ {{\scriptsize ${v_{iX}}(t)$}}}
\put (-109, -12) {{ \scriptsize (d)~{\it t} = 20s}}
\vspace{0em} \caption{Velocity curves of six agents and the formation center at different time.}
\end{center}\vspace{-2em}
\end{figure}

\begin{figure}[!htb]
\begin{center}
\scalebox{0.4}[0.4]{\includegraphics{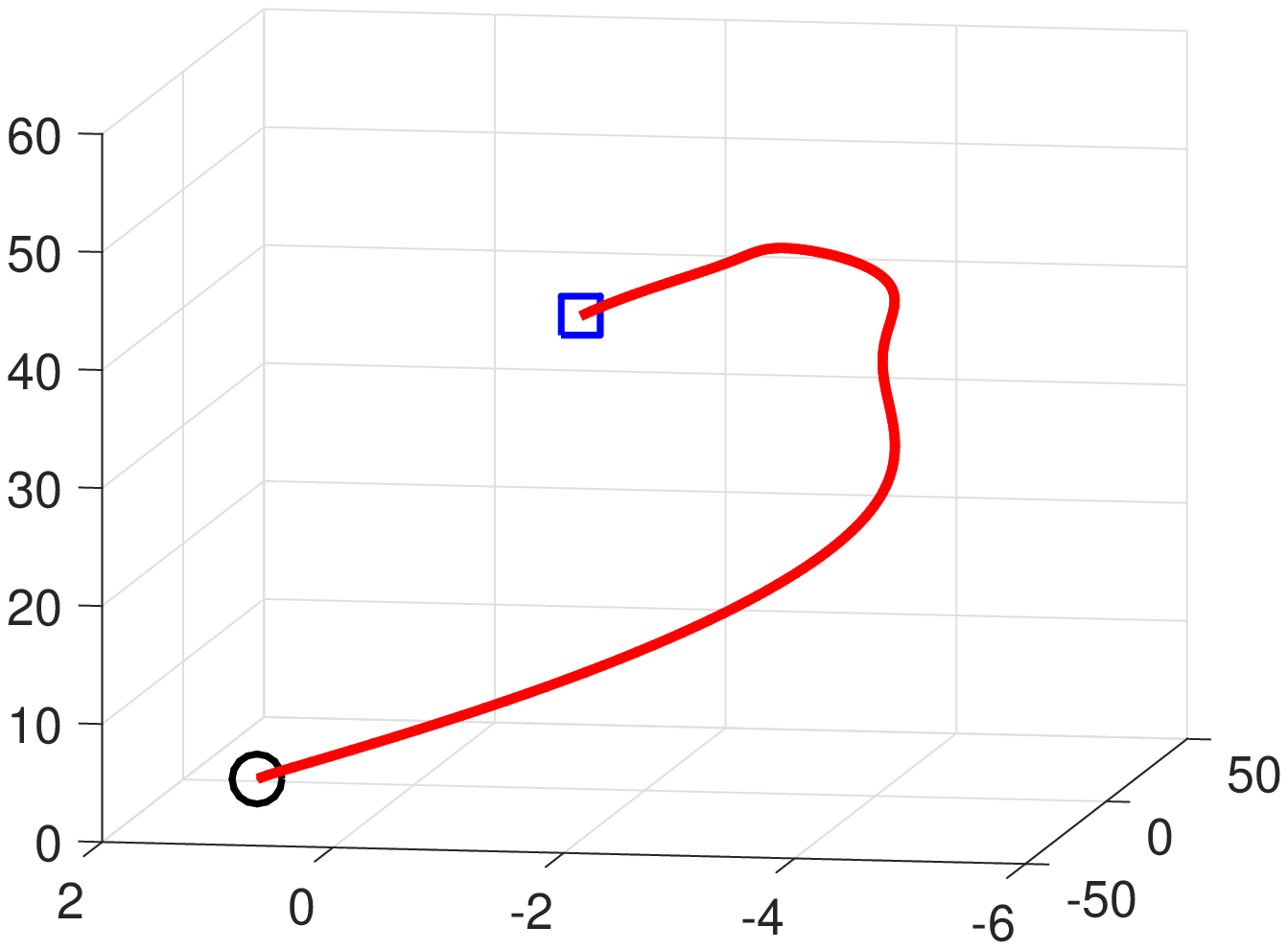}}
\scalebox{0.4}[0.4]{\includegraphics{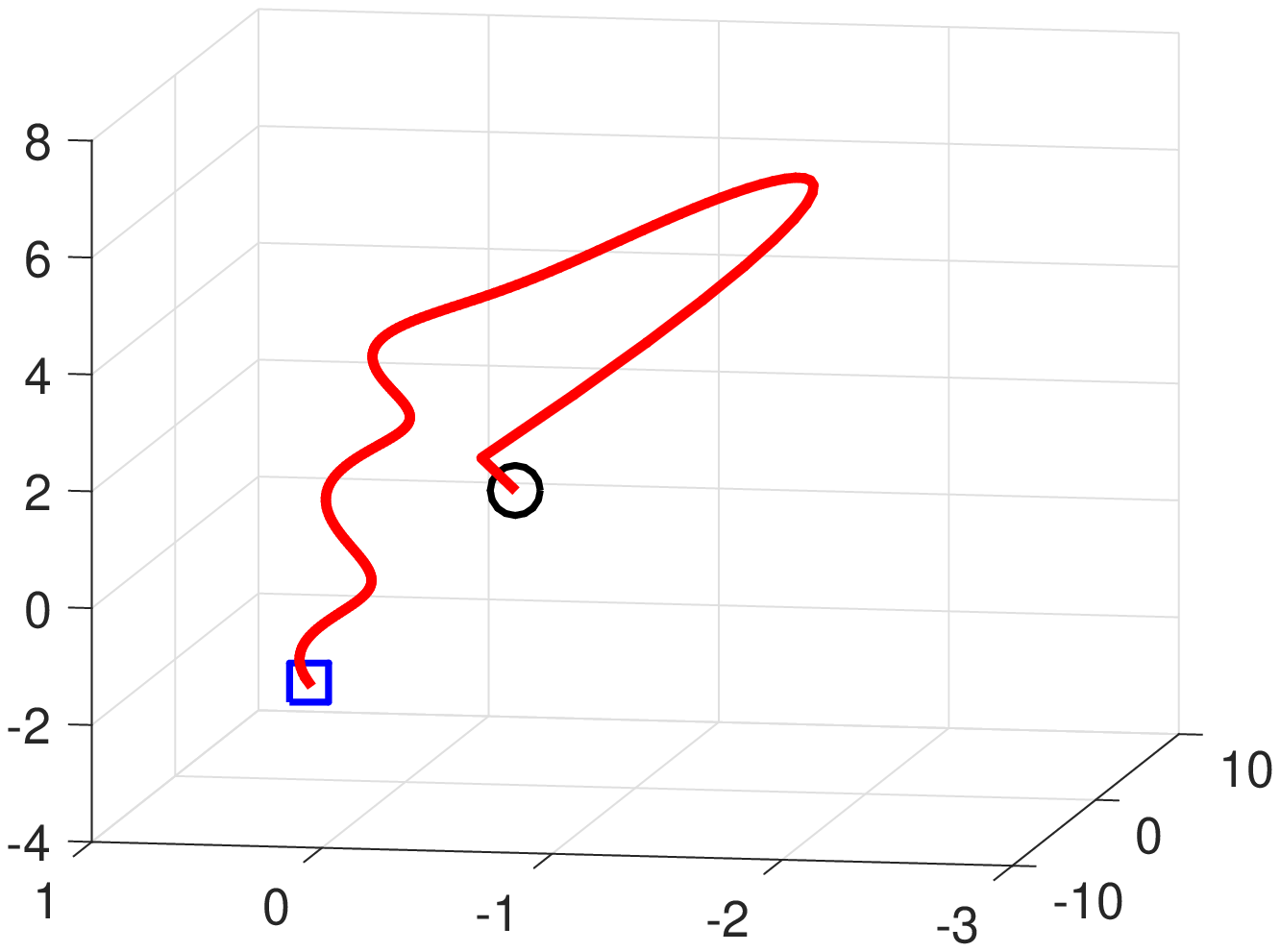}}
\put (-342, 50){\rotatebox{90} {{\scriptsize ${c_{pZ}}(t)$}}}
\put (-283, -5){ {{\scriptsize ${c_{pY}}(t)$}}}
\put (-193, 10){ {{\scriptsize ${c_{pX}}(t)$}}}
\put (-316, -20) {{ \scriptsize (a)~Formation center for positions}}
\put (-168, 50){\rotatebox{90} {{\scriptsize ${c_{vZ}}(t)$}}}
\put (-116, -5){ {{\scriptsize ${c_{vY}}(t)$}}}
\put (-16, 10){ {{\scriptsize ${c_{vX}}(t)$}}}
\put (-149, -20) {{ \scriptsize (b)~Formation center for velocities}}
\vspace{0em} \caption{Curves of the formation center for positions and velocities.}
\end{center}\vspace{-2em}
\end{figure}

\begin{figure}[!htb]
\begin{center}
\scalebox{0.4}[0.4]{\includegraphics{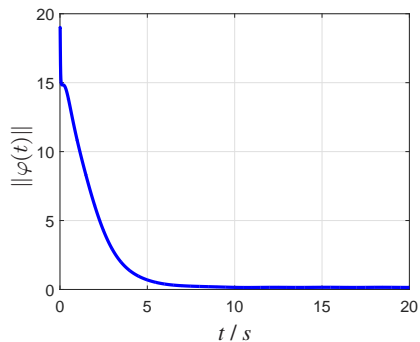}}
\put (-165, 52){\rotatebox{90} {{\scriptsize $\left\| {\varphi (t)} \right\|$}}}
\put (-90, -5){ {{\scriptsize {\it t}~/~\it s}}}
\caption{Trajectory of the time-varying formation error.}
\end{center}\vspace{0em}
\end{figure}

\section{Conclusions}\label{section6}
~~In the current paper, robust time-varying formation design problems for second-order multi-agent systems with external disturbances and directed topologies were studied. A new robust time-varying formation control protocol was proposed with only relative neighboring information and an ESO was designed to estimate and compensate the external disturbances. An explicit expression of the formation center function was derived, where the impacts of the disturbance compensation and the time-varying formation on the motion mode of the whole formation were determined. Sufficient conditions of the robust time-varying formation design were presented via algebraic Riccati equation technique together with the formation feasibility conditions. The tracking performance and the robustness stability of multi-agent systems were analyzed. It was proven that multi-agent systems can reach the expected robust time-varying formation if the gain matrix can be designed and the bandwidth constants of the ESO could be selected properly.

\end{document}